\documentclass{elsart}
\usepackage{epsfig}
\usepackage{amssymb}
\begin{document}
\begin{frontmatter}
\title{PATHWAY MODEL AND NONEXTENSIVE STATISTICAL MECHANICS}
\author{A.M. Mathai}  \ead{mathai@math.mcgill.ca}
\address{Centre for Mathematical Sciences Pala Campus\\
Arunapuram P.O., Pala, Kerala-686574, India\\
and\\
Department of Mathematics and Statistics, McGill University\\
Montreal, Canada H3A2K6}
\author{H.J. Haubold} \ead{hans.haubold@unoosa.org}
\address{Office for Outer Space Affairs, United Nations\\
P.O. Box 500, A1400 Vienna, Austria\\
and\\
Centre for Mathematical Sciences Pala Campus\\
Arunapuram P.O., Pala, Kerala-686574, India}
\author{C. Tsallis}  \ead{tsallis@cbpf.br}
\address{Centro  Brasileiro de Pesquisas Fisicas \\and National Institute of Science and Technology of Complex Systems\\
Rua Xavier Sigaud 150, 22290-180 Rio de Janeiro-RJ, Brazil\\
and\\
Santa Fe Institute\\
1399 Hyde Park Road, Santa Fe, NM 87501, USA}

\begin{abstract}
The established technique of eliminating upper or lower parameters in a general hypergeometric series is profitably exploited to create pathways among confluent hypergeometric functions, binomial functions, Bessel functions, and exponential series. One such pathway, from the mathematical statistics point of view, results in distributions which naturally emerge within nonextensive statistical mechanics and Beck-Cohen superstatistics, as pursued in generalizations of Boltzmann-Gibbs statistics. 
\end{abstract}
\end{frontmatter}
\section{INTRODUCTION}

The celebrated Boltzmann-Gibbs (BG) statistical mechanics, in its classical version, typically holds for many-body systems whose microscopic nonlinear dynamics is {\it ergodic}, either in the entire $\Gamma$ phase space, or in at least one of its subspaces determined by relevant symmetry considerations. For example, for classical ferromagnets (say the infinite-spin Heisenberg ferromagnet in three dimensions) exhibiting a second order phase transition, ergodicity applies to the entire $\Gamma$ phase space for temperatures above the critical one, and only to one of the subspaces generated by the corresponding breakdown of symmetry for temperatures below the critical one. Analogous requirements must be satisfied for quantum systems, where the role of the $\Gamma$ phase space is played by the appropriate Hilbert or Fock spaces. A fundamental question arises for the plethora of physical systems which violate ergodicity in the sense just mentioned: {\it Is it possible to have for them a statistical mechanical theory similar to the usual one, and also connected to thermodynamics?}

It was suggested in 1988 \cite{Tsallis1988} that this is indeed possible based on a
simple hypothesis, namely the generalization of the BG entropy,
given (say in its continuous version) by
\begin{equation}
S_{BG}= -k\int dx \, p(x) \ln p(x) \,.
\end{equation}
The generalization that was then proposed is given by
\begin{equation}
S_{q}= k \frac{1-\int dx \, [p(x)]^q}{q-1} \;\;\;(q \in {\bf R}; \, S_1=S_{BG}).
\end{equation}

Before proceeding, let us mention here that entropic forms
generalizing the Boltzmann-Gibbs-Shannon-von Neumann one have in
fact a long history in information theory, cibernetics and related
areas \cite{entropies,MathaiRathie1975}. Indeed, along the years, the same or similar or related
forms have been introduced again and again as possible mathematical functionals. For example, the Renyi form (defined here below) has been useful as a characterization of multifractal geometry. It appears, however, to be inadequate for thermodynamics since it is not concave for an important range of its parameter $\alpha$, namely for $\alpha>1$, where many physical systems exist. Although independently postulated, the entropic functional $S_q$ turns out to be mathematically very close to those of  Havrda-Charvat, Daroczy, Lindhard-Nielsen, and Mathai-Rathie\footnote{It is by no means rare that abstract ideas formulated in some scientific area eventually find interesting applications in other areas. A celebrated such example is the normal distribution. It was first introduced by Abraham De Moivre in 1733, then by Pierre Simon Laplace in 1774, then by Robert Adrain in 1808, and finally by Carl Friedrich Gauss in 1809, who connected the normal distribution to the theory of errors, applicable in all experimental sciences.}. 
In physics, this type of functional was used \cite{Tsallis1988,CuradoTsallis1991,TsallisMendesPlastino1998} to propose the generalization of the celebrated Boltzmann-Gibbs theory, including the Maxwell-Boltzmann, Fermi-Dirac and Bose-Einstein distributions, as well as their connections to thermodynamics.    

The extremization of $S_q$ under appropriate constraints (nonvanishing first moment, or nonvanishing second moment
if the first moment is zero) yields the $q$-exponential form [$p(x) \propto e_q^{-\beta_1 x}$, or
$p(x) \propto e_q^{-\beta_2\, x^2}$ respectively], where $e_q^{z}\equiv [1+(1-q)z]_+^{1/(1-q)}$, being $[u]_+ = u$
if $u>0$, and zero otherwise; $e_1^z=e^z$. These functions belong to a complex net of related and more general
functions, whose systematic discussion constitutes the aim of the present paper.

The above $q$-exponential functions emerge in a considerable amount
of natural, artificial and social systems. For example (i) The
velocity distribution of (cells of) {\it Hydra viridissima} follows
a $q=3/2$ PDF \cite{UpadhyayaRieuGlazierSawada2001}; (ii) The velocity distribution of (cells of) {\it
Dictyostelium discoideum} follows a $q=5/3$ PDF in the vegetative
state and a $q=2$ PDF in the starved state \cite{Reynolds2010}; (iii) The velocity
distribution in defect turbulence \cite{DanielsBeckBodenschatz2004}; (iv) The velocity
distribution of cold atoms in a dissipative optical lattice \cite{DouglasBergaminiRenzoni2006}; (v)
The velocity distribution during silo drainage \cite{ArevaloGarcimartinMaza2007a,ArevaloGarcimartinMaza2007b}; (vi) The
velocity distribution in a driven-dissipative 2D dusty plasma, with
$q=1.08\pm0.01$ and $q=1.05\pm 0.01$ at temperatures of $30000 \,K$
and $61000\, K$ respectively \cite{LiuGoree2008}; (vii) The spatial (Monte Carlo)
distributions of a trapped $^{136}Ba^+$ ion cooled by various
classical buffer gases at $300\,K$ \cite{DeVoe2009}; (viii) The distributions of
price returns and stock volumes at the stock exchange, as well as the volatility smile  \cite{Borland2002a,Borland2002b,OsorioBorlandTsallis2004,Queiros2005}; (ix) The
distributions of returns of magnetic field fluctuations in the solar
wind plasma as observed in data from  Voyager 1 \cite{BurlagaVinas2005} and from
Voyager 2 \cite{BurlagaNess2009}; (x) The distributions of returns in the Ehrenfest's
dog-flea model \cite{BakarTirnakli2009,BakarTirnakli2010}; (xi)The distributions of returns  in the
coherent noise model \cite{CelikogluTirnakliQueiros2010}; (xii) The distributions of returns of the
avalanche sizes in the self-organized critical
Olami-Feder-Christensen model, as well as in real earthquakes \cite{CarusoPluchinoLatoraVinciguerraRapisarda2007};
(xiii) The distributions of angles in the $HMF$ model \cite{MoyanoAnteneodo2006}; (xiv)
The distribution of stellar rotational velocities in the Pleiades \cite{CarvalhoSilvaNascimentoMedeiros2008}; (xv) The relaxation in various paradigmatic spin-glass
substances through neutron spin echo experiments \cite{PickupCywinskiPappasFaragoFouquet2009}; (xvi) Various
properties directly related with the time dependence of the width of
the ozone layer around the Earth \cite{FerriReynosoPlastino2010}; (xvii) The distribution of
transverse momenta in high energy collisions of electron-positron,
proton-proton, and heavy nuclei (e.g., Pb-Pb and Au-Au) \cite{BediagaCuradoMiranda2000,WilkWlodarczyk2009,BiroPurcselUrmossy2009,CMS1,CMS2,PHENIX,ShaoYiTangChenLiXu2010}, the flux of solar neutrinos \cite{KaniadakisLavagnoQuarati1996}, and the energy distribution of cosmic rays \cite{TsallisAnjosBorges2003}; (xviii)
Various properties for conservative and dissipative nonlinear dynamical systems \cite{LyraTsallis1998,BorgesTsallisAnanosOliveira2002,AnanosTsallis2004,BaldovinRobledo2004,MayoralRobledo2005,PluchinoRapisardaTsallis2007,PluchinoRapisardaTsallis2008,MiritelloPluchinoRapisarda2009,LeoLeoTempesta2010}; (xix) The degree distribution of (asymptotically) scale-free
networks \cite{WhiteKejzarTsallisFarmerWhite2006,ThurnerKyriakopoulosTsallis2007}, and others.

The length of this list illustrates the relevance of a deeper
understanding of the connections of the $q$-exponential functions
with other functions (derivable or not from various entropic forms)
within a variety of pathways, some of which also emerge in applications. This leads us to the next Section.

\section{HYPERGEOMETRIC SERIES}

Consider a confluent hypergeometric series
\begin{equation}
{_1F_1}(a;b;x)=\sum_{r=0}^{\infty}\frac{(a)_r}{(b)_r}\frac{x^r}{r!},~~(c)_r=c(c+1)...(c+r-1), (c)_0=1, c\ne 0.
\end{equation}
If we want to remove the denominator parameter, then a well known technique in the area of
special functions is to replace $x$ by $b\,x$ and then take the limit when $b\rightarrow \infty$. Due to the fact that
\begin{equation}
\lim_{b\rightarrow\infty}\frac{b^r}{(b)_r}=\lim_{b\rightarrow\infty}\frac{b^r}{b(b+1)...(b+r-1)}=1,
\end{equation}
we have
\begin{equation}
\lim_{b\rightarrow\infty}{_1F_1}(a;b;bx)={_1F_0}(a;~ ;x)=(1-x)^{-a},~|x|<1.
\end{equation}
Hence a pathway between the binomial function $(1-x)^{-a}$ and the ${_1F_1}$ series is given by
the limit of $\frac{b^r}{(b)_r}$ when $b\rightarrow\infty$. Going the other way one can build
up a bridge between ${_1F_0}$ and ${_1F_1}$ by introducing $\frac{b^r}{(b)_r}$ into a ${_1F_0}$ series. That is,
\begin{equation}
\sum_{r=0}^{\infty}(a)_r\frac{x^r}{r!}\approx \sum_{r=0}^{\infty}\frac{(a)_r}{(b)_r}\frac{(bx)^r}{r!}={_1F_1}(a;b;bx) ~~~~for\,large~~  b.
\end{equation}
Similarly one can go back and forth from a Bessel function ${_0F_1}$ to a ${_1F_0}$ or to a ${_0F_0}$
which is the exponential series. Let us look at going from a binomial series to an exponential series.
\begin{equation}
(1-x)^{-a}=\sum_{r=0}^{\infty}(a)_r\frac{x^r}{r!}\Rightarrow\lim_{a\rightarrow\infty}\frac{(a)_r}{r!}(\frac{x}{a})^r=\sum_{r=0}^{\infty}\frac{x^r}{r!}={\rm e}^x.
\end{equation}
In other words,
\begin{equation}
{\rm e}^{-cx}=\lim_{a\rightarrow\infty}\sum_{r=0}^{\infty}\frac{(-a)_r}{r!}(\frac{cx}{a})^r=\lim_{a\rightarrow\infty}(1-\frac{cx}{a})^{a}=\lim_{\alpha\rightarrow 1}[1-c(1-\alpha)x]^{\frac{1}{1-\alpha}}.
\end{equation}
Thus a pathway between the exponential function ${\rm e}^{-cx}, c>0$ and the binomial
function $[1-c(1-\alpha)x]^{\frac{1}{1-\alpha}}$ can be created with the help of the pathway parameter $\alpha$. When $\alpha$ is very close to $1$, the binomial and exponential functions are very close to each other and they will be farther apart when $\alpha$ is away from $1$. Observe that ${\rm e}^{-cx}, c>0, 0<x<\infty$ and  $[1-c(1-\alpha)x]^{\frac{1}{1-\alpha}}$, $0<x<\frac{1}{c(1-\alpha)}$, $c>0, \alpha <1$ or $[1+c(\alpha-1)x]^{-\frac{1}{\alpha -1}}$, $c>0, \alpha >1, x>0$ are integrable functions and hence one can create statistical densities out of them. Thus a pathway connecting the three types of densities
\begin{eqnarray}
f_1(x)&=\lambda_1x^{\gamma}{\rm e}^{-cx}, c>0, x>0;\nonumber\\
f_2(x)&=\lambda_2 x^{\gamma}[1-c(1-\alpha)x]^{\frac{1}{1-\alpha}}, c>0,\alpha <1, 0<x<\frac{1}{c(1-\alpha)};\nonumber\\
f_3(x)&=\lambda_3 x^{\gamma}[1+c(\alpha-1)x]^{-\frac{1}{\alpha-1}}, \alpha >1, c>0, x>0,
\end{eqnarray}
where $\lambda_1,\lambda_2,\lambda_3$ are the appropriate normalizing constants, can be created
with the help of the pathway parameter $\alpha$. Observe that in $f_1,f_2$ and $f_3$ one
can replace $x$ by $|x|$, $-\infty<x<\infty$ or $x$ by $|x|^{\delta}, \delta>0$ and still
all the three forms can create densities. Note that $f_1$ stays in the exponential/gamma type
densities, $f_2$ stays as a type-1 beta form and $f_3$ a type-2 beta form. By exploiting these
observations, Mathai  has introduced \cite{Mathai2005} the pathway model connecting exponential type and binomial type functions.\\
\vskip.2cm
Another rich area is the class of Bessel functions. As indicated above, a Bessel function can
be written in terms of a hypergeometric function ${_0F_1}(~ ;b;x)$ and one can remove the
denominator parameter $b$ by replacing $x$ by $bx$ and then using the limit $b\rightarrow\infty$. In other words,
\begin{equation}
\lim_{b\rightarrow\infty}{_0F_1}(~ ;b;-bx)= {_0F_0}(~;~;-x)={\rm e}^{-x}=\lim_{\alpha\rightarrow 1}{_0F_1}(~;\frac{1}{1-\alpha};-\frac{x}{1-\alpha}).
\end{equation}
Thus $\alpha$ can provide a pathway between Bessel functions and exponential functions.
If the exponential form gives the stable situation, then the parameter $\alpha$ will provide a
pathway between stable and chaotic situations. So far this area is not explored. In this connection
one can obtain an interesting result by using the integral representation of a Gauss hypergeometric function ${_2F_1}$, namely,
\begin{eqnarray}
{_2F_1}(a,b;c;-z)=\frac{\Gamma(c)}{\Gamma(a)\Gamma(c-a)}&\int_0^1 x^{a-1}(1-x)^{c-a-1}(1+zx)^{-b}{\rm d}x,\\
&\Re(a)>0, \Re(c-a)>0, |z|<1.\nonumber
\end{eqnarray}
Hence,
\begin{eqnarray}
{_1F_1}(a;c;-z)&=&\lim_{b\rightarrow\infty}{_2F_1}(a,b;c;-\frac{x}{b}), |x|<1\nonumber\\
&=&\frac{\Gamma(c)}{\Gamma(a)\Gamma(c-a)}\lim_{b\rightarrow\infty}\int_0^1 x^{a-1}(1-x)^{c-a-1}(1+\frac{zx}{b})^{-b}{\rm d}x, |z|<1 \nonumber \\
&=&\frac{\Gamma(c)}{\Gamma(a)\Gamma(c-a)}\int_0^1 x^{a-1}(1-x)^{c-a-1}{\rm e}^{-zx}{\rm d}x, |z|<1.
\label{eq2}
\end{eqnarray}
Thus a pathway between ${_1F_1}$ and ${_2F_1}$ is given by (\ref{eq2}). Many such results can be
obtained by using this technique of eliminating one or more numerator or denominator parameters from a general hypergeometric series ${_pF_q}$.\\
\vskip.2cm
Thus for a real scalar random variable $x$, the pathway density can be written in the following form:
\begin{equation}
f_4(x)=\lambda_4|x|^{\gamma}[1-a(1-\alpha)|x|]^{\frac{1}{1-\alpha}}, a>0,1>a(1-\alpha)|x|,\alpha<2.
\label{general}
\end{equation}
A more general form of the pathway density is the following:
\begin{equation}
f_5(x)=\lambda_5|x|^{\gamma}[1-a(1-\alpha)|x|^{\delta}]^{\frac{\eta}{1-\alpha}},
\label{moregeneral}
\end{equation}
where $a>0$ and $(\eta,\delta,\gamma,\alpha)$ are such that $f_5(x)$ is normalizable.
A large number of commonly used
statistical densities can be seen to be particular cases of (\ref{moregeneral}), details may be seen in \cite{Mathai2005,MathaiHaubold2007a,MathaiHaubold2007b}.
From the point of view of mathematical statistics, nonextensive statistics \cite{Tsallis1988,CuradoTsallis1991,TsallisMendesPlastino1998,GellMannTsallis2004,Tsallis2004,Tsallis2009} with constant density of states is a particular
case of (\ref{general}) for $\gamma=0, x>0$. The case $\gamma \ne 0$ can be seen as the particular case when the density of states is given by a power law (which is quite frequent in many physical systems). One of the forms of the Beck-Cohen superstatistics \cite{BeckCohen2003,Beck2006} is
a special case of (\ref{general}) for $\gamma=0, \alpha >1, x>0$.\\

\section{DENSITY FROM OPTIMIZATION OF ENTROPY}

In situations when an appropriate density is selected, one guiding principle is the maximization of entropy.
Entropy or a measure of uncertainty in a scheme or ``information'' in a scheme is traditionally measured
by Shannon entropy. Consider a discrete distribution $P'=(p_1,...,p_k), p_i>0, i=1,...,k, p_1+...+p_k=1$.
This may also be looked upon as the sample space or the sure event $S$ is partitioned into mutually exclusive
and totally exhaustive events $A_1,...,A_k, A_1\cup ...\cup A_k=S, A_i\cap A_j=O$ for all $i$ and $j$, $i\ne j$
with the probability of the event $A_i$, denoted by $p_i=Pr(A_i), i=1,...,k$. If any $p_i$ is allowed to
take the value zero also, then $p_i\ge 0, i=1,...,k$. Shannon entropy on this scheme is $S(P)$, where
\begin{equation}
S(P)=-\sum_{i=1}^k p_i\ln p_i.
\end{equation}
When a $p_i=0$, $p_i\ln p_i$ is to be interpreted as zero. Several
characterization theorems on $S(P)$ or axiomatic definitions may be
seen from [3]. There are several extensions or generalizations of
the measure $S_k(P)$. Classical generalizations in information theory are the Havrda-Charv\'at measure $H_{k,\alpha}(P)$, and the
R\'enyi measure $R_{k,\alpha}(P)$, where
\begin{equation}
H_{\alpha}(P)=\frac{\sum_{i=1}^k p_i^{\alpha}-1}{2^{1-\alpha}-1}\, , R_{\alpha}(P)=\frac{\ln(\sum_{i=1}^k p_i^{\alpha})}{1-\alpha},\alpha\ne 1,\alpha >0.
\end{equation}
These are generalizations in the sense that when $\alpha\rightarrow 1, H_{\alpha}(P)\rightarrow S(P)$
and $R_{\alpha}(P)\rightarrow S(P)$. Out of these, $S(P)$ and $R_{\alpha}(P)$ are additive and
$H_{\alpha}(P)$ is nonadditive. The additivity property is defined as follows: Consider a bivariate
discrete distribution in the sense $p_{ij}>0, i=1,...,m, j=1,...,n$ such that $\sum_{i=1}^m\sum_{j=1}^n p_{ij}=1$.
What happens if there is the product probability property (PPP), which in statistical literature is known as
statistical independence. 
What happens is 
that
$p_{ij}=p_iq_j, p_1+..+p_m=1, q_1+...+q_n=1$ or there is the product probability property. When PPP
holds, if the entropy in the joint distribution $(P,Q)=(p_{ij}),i=1,...,m, j=1,...,n$ is the sum of
the entropies on $P$ and $Q$ then we say that there is additivity. It is easily seen that there is
additivity in $S(P)$ and $R_{\alpha}(P)$, that is,
\begin{equation}
R_{\alpha}(P,Q)=R_{\alpha}(P)+R_{\alpha}(Q) ~~  and ~~ S(P,Q)=S(P)+S(Q).
\end{equation}

This additivity holds due to the logarithmic nature of the function in $S_k(P)$ and $R_{\alpha}(P)$ and
the logarithm of a product of positive quantities being the sum of the logarithms. It is explained in \cite{MathaiHaubold2007a}
 that logarithmic function enters into an entropy measure due to the recursivity axiom which leads into a logarithmic function necessarily.\\

In the following we will concentrate on the $q$-type of
generalization of entropy measures, and review, for completeness, how the extremization of generalized entropies yields the probability density which correspond to stationary states. It was postulated [1]  the
entropy
\begin{equation}
S_{\alpha}(P)=\frac{\sum_{i=1}^k p_i^{\alpha}-1}{1-\alpha}, \alpha\ne 1, \alpha >0.
\end{equation}
To avoid confusion, let us mention that, in most of the literature of nonextensive statistical mechanics,
the index $\alpha$ is noted $q$, and the entropy  $S_{\alpha}$ is noted $S_q$. The Shannon form is obtained as the $q \equiv \alpha \to 1$ limit.

The normalizing factor in Havrda-Charv\'at entropy $H_{\alpha}(P)$, namely $(2^{1-\alpha}-1)$,
is replaced by $(1-\alpha)$. In the continuous case, the nonadditive entropy upon which nonextensive
statistical mechanics is built is then,
\begin{equation}
S_{\alpha}(f)=\frac{\int_{x}[f(x)]^{\alpha}{\rm d}x-1}{1-\alpha}, \alpha\ne 1, \alpha >0.
\label{functional}
\end{equation}

Over all functions $f$, what is that particular $f$ which will
optimize the nonadditive entropy in (\ref{functional})? If calculus of variation
principle is used, then the Euler equation for optimizing the entropy
$S_{\alpha}$ under the restrictions
\begin{equation}
\int_xf(x){\rm d}x=1 ~~  and  ~~ \int_xxf(x){\rm d}x=E(x)= ~~ fixed~~ , f(x)\ge 0 \,,   \forall  x
\end{equation}
will yield the equation,
\begin{eqnarray}
\frac{\partial}{\partial f}[f^{\alpha}-\lambda_1f-\lambda_2xf]=0&\Rightarrow f=[\lambda+\lambda_2x]^{\frac{1}{\alpha-1}}\nonumber\\
&=\lambda[1+(\alpha-1)x]^{\frac{1}{\alpha-1}}
\label{21}
\end{eqnarray}
by taking $\frac{\lambda_2}{\lambda_1}=\alpha-1$ and
$\lambda_1^{\frac{1}{\alpha-1}}=\lambda$, where $\lambda_1$ and
$\lambda_2$ are Lagrangian multipliers. The quantity $\lambda$ can act as the
normalizing constant. The condition $E(x)=$ fixed, where $E$ denotes
the expected value, can be interpreted as the principle of
conservation of the quantity $x$. When $\alpha\rightarrow 1,
f=\lambda {\rm e}^{-x}$ which is an exponential function. The
derivation in (\ref{21}) does not yield nonextensive statistics in its most convenient form.
But (\ref{21}) gives an exponential function when $\alpha\rightarrow 1$
and this exponential function is directly related to what is known in the literature as the
$q$-exponential function. In order to circumvent some difficulties,
it was replaced (\cite{Tsallis1988,TsallisMendesPlastino1998})  the second condition that $E(x)$ is
fixed by fixing the expected value in the escort distribution. The
escort density is given by
\begin{equation}
g(x)=\frac{f^{\alpha}(x)}{\int_xf^{\alpha}(x){\rm d}x}
\end{equation}
and then nonextensive statistics has the form
\begin{equation}
f=\lambda[1-(1-\alpha)x]^{\frac{1}{1-\alpha}}.
\end{equation}
This form can produce densities for $\alpha <1, \alpha >1$ and
$\alpha\rightarrow 1$ and further, this form satisfies the power-law differential equation
\begin{equation}
\frac{{\rm d}}{{\rm d}x}(\frac{f}{\lambda})=-(\frac{f}{\lambda})^{\alpha}.
\end{equation}
One can introduce a general measure of entropy, which in the discrete
and continuous cases are denoted by $M_{\alpha}(P)$ and $M_{\alpha}(f)$ respectively, where
\begin{equation}
M_{\alpha}(P)=\frac{\sum_{i=1}^k p_i^{2-\alpha}-1}{\alpha -1},\alpha\ne 1,\alpha<2,\, M_{\alpha}(f)=\frac{\int_x[f(x)]^{2-\alpha}{\rm d}x-1}{\alpha-1},
\label{entropy}
\end{equation}
$\alpha\ne 1,\alpha<2$. A characterization of $M_{\alpha}(P)$ is
given in \cite{MathaiRathie1975} (see also \cite{MathaiHaubold2007a,MathaiHaubold2007b,FerriMartinezPlastino2005}). If $M_{\alpha}(f)$ is optimized under the conditions
that $E(x)=$ fixed and that $f(x)$ is a density, then the Euler
equation becomes
\begin{equation}
\frac{\partial}{\partial f}[f^{2-\alpha}-\lambda_1f+\lambda_2 xf]=0\Rightarrow f=\lambda[1-a(1-\alpha)x]^{\frac{1}{1-\alpha}},
\label{26}
\end{equation}
where $a>0, 1-a(1-\alpha)x>0, \frac{\lambda_2}{\lambda_1}$ is taken
as $a(1-\alpha)$ with $a>0$ and
$\left(\frac{\lambda_1}{2-\alpha}\right)^{\frac{1}{1-\alpha}}$ is
taken as $\lambda$. Observe that (\ref{26}) readily gives densities for
$\alpha <1, \alpha >1$ and $\alpha \rightarrow 1$. Further, the
entropy itself can be expressed as
\begin{equation}
M_{\alpha}(P)=\frac{E(p^{1-\alpha})-1}{\alpha-1},  M_{\alpha}(f)=\frac{E[f^{1-\alpha}(x)]-1}{\alpha-1}
\end{equation}
where $(1-\alpha)$ can be interpreted as the strength of information
in $f$ and this expected value is also associated with Kerridge's
``inaccuracy'' measure. As a simple mathematical remark, let us mention that if the entropy in (\ref{entropy}) is optimized in an {\it ad hoc} manner, namely that for all
$f(x)$ such that $f(x)\ge 0$ for all $x>0$, $\int_xf(x){\rm
d}x<\infty, \int_x[x^{\gamma(1-\alpha)}] f(x){\rm d}x=$ fixed  and
$\int_x[x^{\gamma(1-\alpha)+\delta}]f(x){\rm d}x=$ fixed, then we end
up with the density
\begin{equation}
f(x)=\lambda x^{\gamma}[1-a(1-\alpha)x^{\delta}]^{\frac{1}{1-\alpha}}, -\infty <\alpha <2, a>0,
\label{lambda}
\end{equation}
and $\lambda$ is the normalizing constant. Through trivial changes in the notation, this expression recovers that of (\ref{moregeneral}).

As already mentioned, for $\gamma=0,\delta=1$ in (\ref{lambda}) one has a particular case of nonextensive statistics. For $a>0, \alpha >1$ in (\ref{lambda}) one has a particular case of
the superstatistics of Beck and Cohen \cite{BeckCohen2003,Beck2006}. For $\alpha <1$, (\ref{lambda}) gives a generalized
type-1 beta form for $0<x^{\delta}<\frac{1}{a(1-\alpha)}$, and for $\alpha >1$, (\ref{lambda}) gives
 a generalized type-2 beta form. Superstatistics can produce only the type-2 beta form and
 not the type-1 beta form.

\section{FINAL REMARKS}

We utilize the established technique of eliminating upper or lower parameters in a general hypergeometric series to create pathways among confluent hypergeometric functions, binomial functions, Bessel functions, and exponential series. Mathai's pathway, from the mathematical statistics point of view, results in distributions which also emerge within nonextensive statistics and Beck-Cohen superstatistics, pursued as generalizations of Boltzmann-Gibbs statistics. It was shown that this pathway model can also be derived by optimizing a generalized entropic measure. Through Mathai's pathway approach, exponential and binomial type functions are connected through the pathway model parameter. The same pathway model also leads to a link between Bessel functions and exponential functions. The pathway model covers statistical densities emanating in nonextensive statistics and Beck-Cohen superstatistics as special cases of (\ref{lambda}). Related results are obtained by optimizing a general measure of entropy in (\ref{entropy}) (see also \cite{Tsallis1988,TsallisMendesPlastino1998,FerriMartinezPlastino2005}). An open problem is identified that would allow to entropically derive a general density of the form (\ref{lambda}) within physically meaningful circumstances. Summarizing, relations between Mathai's pathway model and nonextensive statistics and Beck-Cohen superstatistics were exhibited.

\section*{Acknowledgments}

A.M.M. and H.J.H. would like to thank the Department of Science and Technology, Government of India for the financial
assistance for this work under grant No. SR/S4/MS:287/05. C.T. acknowledges partial support from CNPq and Faperj (Brazilian Agencies).

\end{document}